\documentclass[conference,letter]{IEEEtran}
\usepackage[utf8]{inputenc}
\usepackage[english]{babel}
\usepackage{graphicx}
\usepackage{xcolor}
\usepackage{cite}
\usepackage[T1]{fontenc} 
\usepackage{amssymb}
\usepackage{amsfonts}
\usepackage{booktabs}
\usepackage{booktabs}
\usepackage{amsmath}
\usepackage[mathscr]{euscript}
\usepackage{multirow}
\usepackage{soul}
\usepackage{mathtools}
\usepackage[acronym]{glossaries}
\usepackage{amsmath}
\usepackage{siunitx}
\usepackage{multirow}
\usepackage{algorithm}
\usepackage{algorithmic}
\usepackage{wrapfig}
\usepackage{lipsum}

\IEEEoverridecommandlockouts

\newcommand{\sps}[1]{\textcolor{magenta}{[#1]}}

\DeclareSIUnit{\dBm}{dBm}

\newcommand{\TransmittedData}{d^j_{i,t}}
\newcommand{\BatteryStateDynamics}{b^j_t}
\newcommand{\QueueStateDynamics}{q^j_{i,t}}
\newcommand{\PowerConversionEfficiency}{\lambda^j}
\newcommand{\HarvestedPower}{\tilde{p}_t}
\newcommand{\EHChannelScaleParameter}{\tilde{\zeta}^j}

\newcommand{\EHChannelNorm}{\tilde{h}^j_t}
\newcommand{\TransmissionPower}{p^j_{i,t}}
\newcommand{\PrimaryMaxPower}{P_0}
\newcommand{\SecondaryMaxPower}{P_1}
\newcommand{\PrimaryTransmissionPower}{p^j_{0,t}}

\newcommand{\CommunicationTime}{\alpha^j_{i,t}}
\newcommand{\PrimaryCommunicationTime}{\alpha^j_{0,t}}
\newcommand{\Rayleighdistribution}{f(\zeta_i)}

\newcommand{\SecondaryCommunicationTime}{\alpha^j_{1,t}}
\newcommand{\EHTime}{\tilde{\alpha}^j_t}
\newcommand{\tx}{\text{TX}}
\newcommand{\rx}{\text{RX}}
\newcommand{\txPrimary}{\tx_0}
\newcommand{\txSecondary}{\tx_1}
\newcommand{\rxPrimary}{\rx_0}
\newcommand{\rxSecondary}{\rx_1}
\newcommand{\primary}{(\txPrimary,\rxPrimary)}
\newcommand{\secondary}{(\txSecondary,\rxSecondary)}
\newcommand{\txI}{\tx_i}
\newcommand{\rxI}{\rx_i}
\newcommand{\systemI}{(\txI,\rxI)}
\newcommand*{\starnr}{\stepcounter{equation}\tag{\theequation}}


\usepackage{titlesec}
\let\subparagraph\relax
\titlespacing{\section}{0pt}{5pt plus 2pt minus 1pt}{4pt plus 1pt minus 1pt} 
\titlespacing{\subsection}{0pt}{4pt plus 2pt minus 1pt}{2pt plus 1pt minus 1pt} 
\begin{document}



\newacronym{EH}{EH}{energy harvesting} 
\newacronym{AWGN}{AWGN}{additive white Gaussian noise} 
\newacronym{WSN}{WSN}{wireless sensor network} 
\newacronym{MDP}{MDP}{markov decision process} 
\newacronym{L2RL}{L2RL}{lifelong reinforcement learning} 
\newacronym{RL}{RL}{reinforcement learning} 
\newacronym{UAV}{UAV}{unmanned aerial vehicle} 
\newacronym{AoI}{AoI}{age of information} 
\newacronym{IoT}{IoT}{internet of things} 
\newacronym{MT-L2RL}{MT-L2RL}{multi-tasks L2RL} 
\newacronym{SWIPT}{SWIPT}{simultaneous wireless information and power transfer} 
\newacronym{PPP}{PPP}{poisson point process} 
\newacronym{TX}{TX}{transmitter}
\newacronym{RX}{RX}{receiver}
\newacronym{ML}{ML}{machine learning} 
\newacronym{PDF}{PDF}{probability density function}
\newacronym{KB}{KB}{knowledge base}
\newacronym{QoS}{QoS}{quality-of-service}

\title{Multi-Task Lifelong Reinforcement Learning for Wireless Sensor Networks}
\author{
\IEEEauthorblockN{
Hossein~Mohammadi~Firouzjaei\IEEEauthorrefmark{1}, 
Rafaela~Scaciota\IEEEauthorrefmark{1}\IEEEauthorrefmark{2}, 
and Sumudu~Samarakoon\IEEEauthorrefmark{1}\IEEEauthorrefmark{2}
}
\IEEEauthorblockA{
	\small%
	\IEEEauthorrefmark{1}%
	Centre for Wireless Communication, University of Oulu, Finland \\
    \IEEEauthorrefmark{2}%
    Infotech Oulu, University of Oulu, Finland \\
	Email: \{hossein.mohammadifirouzjaei, rafaela.scaciotatimoesdasilva, sumudu.samarakoon\}@oulu.fi 
}
\vspace{-20pt}
\thanks{
This work was supported by the projects NSF-AKA CRUISE (GA 24304406), Infotech-R2D2, and 6G Flagship (Grant Number 369116) funded by the Research Council of Finland.}
}

\maketitle

\begin{abstract} 
Enhancing the sustainability and efficiency of \glspl{WSN} in dynamic and unpredictable environments requires adaptive communication and \gls{EH} strategies.
We propose a novel adaptive control strategy for \glspl{WSN} that optimizes data transmission and \gls{EH} to minimize overall energy consumption while ensuring queue stability and energy storing constraints under dynamic 
environmental conditions.
The notion of adaptability therein is achieved by transferring the known environment-specific knowledge to new conditions resorting to the \gls{L2RL} concepts.
We evaluate our proposed method against two baseline frameworks: Lyapunov-based optimization, and policy-gradient \gls{RL}. 
Simulation results demonstrate that our approach rapidly adapts to changing environmental conditions by leveraging transferable knowledge, achieving near-optimal performance approximately $30\%$ faster than the \gls{RL} method and $60\%$ faster than the Lyapunov-based approach.
The implementation is available at our GitHub repository for reproducibility purposes\cite{firouzjaei2025github}.
\end{abstract}

\begin{IEEEkeywords}
Wireless Sensor Networks, Energy Harvesting, Lifelong Reinforcement Learning
\end{IEEEkeywords}
\glsresetall

\section{Introduction}

\Gls{EH} has emerged as a promising solution to improving the performance of \glspl{WSN}, which are often constrained by the finite capacity of traditional battery-powered nodes~\cite{Zhang.18}. 
However, despite the benefits of \gls{EH}, the limited processing power and limited data storage capabilities of these nodes remain as significant challenges, especially in the context of large-scale deployments~\cite{Lall.16}.
Such limitations result in delays and energy losses in \glspl{WSN} due to the accumulation of energy and data in their respective buffers. 
Furthermore, the dynamical nature of the environments leads to non-stationary data distributions, making data modeling and forecasting challenging~\cite{O'Reilly.14}. 
As a result, implementing adaptable transmission and \gls{EH} strategies become essential towards designing energy efficiency and latency sensitive \glspl{WSN}. 

Recent advances in \glspl{WSN} have presented research interest in overcoming battery limitations through \gls{EH}. In~\cite{Blasco.13}, a Q-learning-based approach is proposed to optimize energy management in wireless \gls{EH} communication systems under stochastic data and energy arrivals; however, it assumes stationary \glspl{MDP} for both, which may not hold in real-world, evolving conditions. Similarly,\cite{Ortiz.16} studies a point-to-point EH communication scenario where only causal information about the EH process and the channel is available at the transmitter. They model the system as a \gls{MDP} and apply \gls{RL} with linear function approximation to derive a power allocation policy that maximizes throughput, but this method also relies on the assumption of stationary statistical properties for the EH process. Addressing more complex scenarios,\cite{Zhang.19} presents an \gls{RL}-based algorithm that learns from historical statistics to optimize EH and resource allocation policies while satisfying heterogeneous, statistically delay-bounded \gls{QoS} constraints in 5G wireless ad-hoc networks. Nonetheless, this approach still falls short in accounting for the high variability and unpredictability of the EH process, where energy sources may switch on and off randomly, further complicating the maintenance of policy optimality in dynamic environments.

To address the challenges of non-stationary environments, recent research has focused on improving standard \gls{RL} frameworks to enable adaptive learning in dynamic conditions. 
For example, in~\cite{Padakandla.20} a detection algorithm is introduced that monitors and reacts to variations in environmental statistics, allowing the agent in \gls{RL} to adjust policies and maintain optimal performance dynamically. 
The proposed approach improves decision-making in environments characterized by unpredictable changes, such as traffic signal control and sensor energy management, by integrating change detection with Q-learning. Building on this, the authors in~\cite{Gong.24} advances the concepts by proposing a lifelong learning framework. The authors solution utilizes \gls{UAV}s as mobile agents to optimize energy consumption and \gls{AoI} in non-stationary \gls{IoT}  networks.
Although the authors in~\cite{Gong.24} considers dynamic arrival rates and device-specific characteristics, it does not address the evolving conditions of communication and \gls{EH} channels. Focusing on adaptable solutions for dynamic environments, the authors in~\cite{Wu.23}, propose a lifelong learning \gls{RL}-based algorithm for online computing resource allocation in \gls{IoT} devices, \glspl{UAV}, and satellites, aiming to optimize the trade-off between \gls{AoI} and energy consumption.  While this approach is promising, it does not fully address the challenges of non-stationary environments, where dynamic topology and fluctuating network traffic require algorithms that can continuously adapt to new tasks and environmental changes.

The main contribution of this paper is the introduction of a novel adaptable \gls{L2RL}-based algorithm for joint \gls{EH} and communication in non-stationary \glspl{WSN}. 
For simplicity, we assume a \gls{WSN} composed of two subsystems, each with its own transmitter–receiver pair. 
The primary subsystem operates with a stable energy source and supports both data transmission and wireless energy transfer, while the secondary subsystem uses the same communication technology but relies on harvested energy from the primary.
Therein, each communication system must process data in an energy-efficient manner while coping with dynamic \gls{EH} conditions, limited resource constraints on transmission, storage, and queue stability requirements.
First, we pose this as a series of optimization problems, each defined for specific environmental conditions, but sharing common state and action spaces.
Hence, we define solving an optimization problem with specific environmental conditions as a single \emph{task} inline with the \gls{L2RL} framework. 
Then, we cast the above set of problems as a \gls{MDP} that solves multiple tasks. 
Since the state and action spaces remain unchanged, while the network operates under non-stationary \gls{EH} conditions, the problem can be framed as a multi-task scenario in the context of \gls{L2RL}. Each task is defined by different \gls{EH} dynamics, but all share the same state and action spaces.
We evaluate the proposed \gls{MT-L2RL} algorithm against two well-established baseline methods: i) a traditional \gls{RL} framework that learns optimal policies through interaction with the environment, and ii) a Lyapunov optimization approach that applies Lyapunov drift theory to ensure system stability and optimize performance simultaneously.
The simulation results demonstrate that \gls{MT-L2RL} consistently outperforms both baselines across dynamic environments, converging to near-optimal performance approximately $30\%$ to $60\%$ faster. This accelerated convergence reflects the system's strong adaptability, enabled by leveraging transferable knowledge from previous tasks to quickly adjust to new tasks.

The remainder of the paper is structured as follows. 
Section~\ref{sec:system_model} outlines the system model and the problem formulation. The \gls{L2RL}-based proposed solution is introduced in Section~\ref{sec:L2RL_approach}.
Section~\ref{sec:simulation_results} evaluates and compares the solution with the baseline methods. 
Finally, Section~\ref{sec:conclusion} provides the conclusions.

\noindent
\textit{Notation:} 
Vectors are in lowercase and bold. 
Matrices are uppercase bold.
$\mathbf{I}$ refers to the identity matrix. 
$\text{vec}(\mathbf{A})$ refers to vectorization of $\mathbf{A}$. 
$\mathbb{R}$ refers to the set of real numbers. 

\section{System Model \& Problem Formulation} \label{sec:system_model}

\begin{figure}
    \centering
    \includegraphics[width=8.8cm]{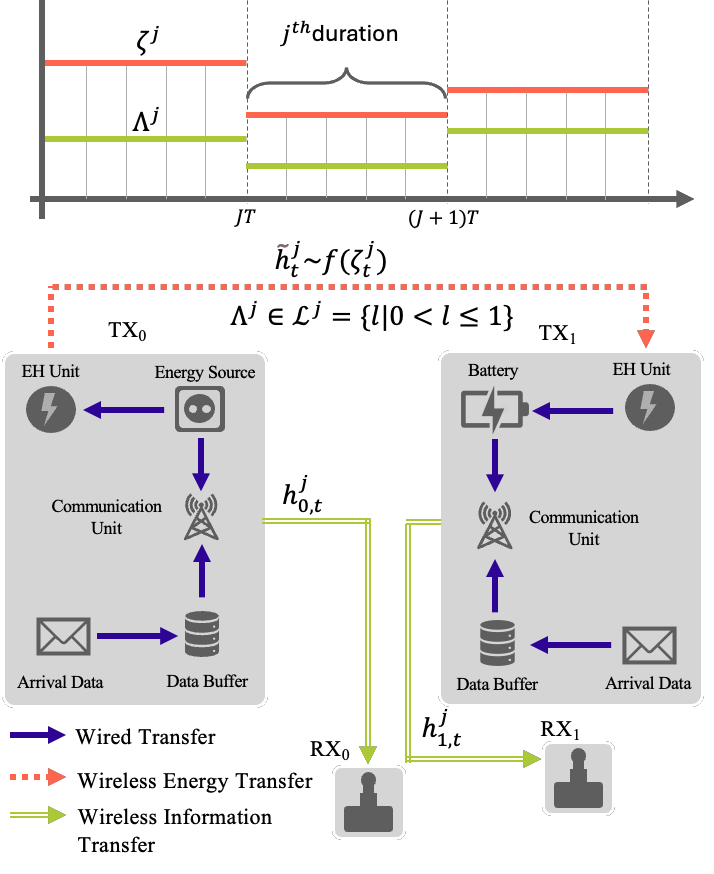}
    \caption{System model comprising wireless sensors, $\primary$ and $\secondary$, in a non-stationary environment. 
    }
    \label{fig:system_model}
\end{figure}

We consider a \gls{WSN} consisting of primary and secondary sensor nodes, each equipped with a single-antenna \gls{TX} for data transmission and a single-antenna \gls{RX} for control signals, denoted by $\primary$ and $\secondary$, respectively, as illustrated in Fig.~\ref{fig:system_model}. 
In the primary system, $\txPrimary$ is connected to a stable energy source and carries out data transmissions to $\rxPrimary$ while simultaneously acting as a power beacon for the secondary system, i.e., it is capable of \gls{SWIPT}.
$\txSecondary$ is equipped with a rechargeable battery and harvests energy from $\txPrimary$ over a wireless \gls{EH} channel.
The communication between $\secondary$ is carried out with the harvested energy.

We assume both primary and secondary communication systems follow queuing models under channel dynamics and limited communication resources.
As such, the queue state $q_{i,t}$ dynamics of $\txI$ with $i\in\{0,1\}$ at a time slot $t$ is given by~\cite{Qiu.18},
%
\begin{equation}\label{eqn:queue_dynamics}
    q_{i,t+1} = \max \{ q_{i,t} - d_{i,t}, 0 \} + a_{i,t},
\end{equation}
where $d_{i,t}$ is the transmitted data and $a_{i,t}$ is the arrival data that follows a homogeneous \gls{PPP}.
It is worth highlighting that a fraction $\alpha_{i,t}$ of a time slot is used for the data communication, in which, the transmitted data at $t$ follows
%
\begin{equation}\label{eqn:transmitted_data}
    d_{i,t} = W \log_2 \left(1 + \frac{p_{i,t} h_{i,t}}{N_0}\right) \alpha_{i,t},
\end{equation}
where $W$ is the transmission bandwidth, $N_0$ denotes the power spectral density of \gls{AWGN}, and $p_{i,t}$ is the transmission power of $\tx_i$ at time slot $t$. 
Here, $h_{i,t} \sim \Rayleighdistribution$ represents the norm of the Rayleigh fading channel gain between the $\systemI$ pair, modeled with a scale parameter $\zeta_i > 0$. 

Let $\tilde{h}_t$ be the channel gain between $\txPrimary$ and $\txSecondary$ during the \gls{EH}.
We assume the \gls{EH} channel follows a Rayleigh fading model parameterized by $\tilde{\zeta}$. 
Given that $\txPrimary$ uses $p_{0,t}$ of its maximum available power $\PrimaryMaxPower$ for data transmission, the remaining power, $P_0 - p_{0,t}$, is allocated to energy transfer. This reflects a system constraint: the total power used for both data transmission and energy harvesting (EH) must not exceed $P_0$. This setup highlights a fundamental trade-off between communication performance and energy harvesting. Allocating more power to data transmission reduces the power available for \gls{EH}, and vice versa.

Moreover, the amount of energy harvested depends on the \gls{EH} channel conditions, captured by $\tilde{h}_t$, which represents the instantaneous channel gain for energy transfer. The instantaneous harvested power at $\txSecondary$ is given by~\cite{Wang.21}:
\begin{equation}\label{eqn:harvested_power}
    \HarvestedPower = \lambda \big(\PrimaryMaxPower - p_{0,t}\big) \tilde{h}_t,
\end{equation}
where $\lambda \in \mathscr{L} = \{l \mid 0 < l \leq 1\}$ is the power conversion efficiency. 
At $\txSecondary$, the harvested energy over a fraction $\tilde{\alpha}_t$ is stored in a battery.
It is assumed that $p_{1,t} = \SecondaryMaxPower$ , where $\SecondaryMaxPower$ denotes $\txSecondary$ maximum transmission power.
In this view, battery state $b_t$ dynamics corresponding to energy arrival, storage, and consumption is modeled as
\begin{equation}\label{eqn:battery_queue_dynamics}
    b_{t+1} = \max \{b_t - p_{1,t} \alpha_{1,t},0 \} + \HarvestedPower \tilde{\alpha}_t.
\end{equation}

We assume that the system operates in a non-stationary environment where the \gls{EH} conditions change after a certain period of time. 
Specifically, the power conversion efficiency $\lambda$ and the scale parameter $\tilde{\zeta}$, which characterize the \gls{EH} process, are subject to change over certain time intervals.
In this framework, each \emph{task} corresponds to an interval where the system must be designed to adapt its strategy to the specific \gls{EH} conditions of that task.
To represent the system behavior under such changes, we introduce a superscript $j$ with the system variables and parameters to indicate their dependence on task $j$. 
That is, within a task $j$, the \gls{EH} conditions are fixed and given by $\lambda^j$ and $\tilde{\zeta}^j$, which are selected from a predefined set.
These profiles may represent different environmental contexts.  
Hence, the overall system operation can be designed as a multi-task design problem.
We define stationary periods of length \( T \) as follows
\begin{equation}
    \mathcal{T}^j = \{ j\times T, (j\times T)+1, \dots, (j\times T) + T - 1 \}, \quad j \in \mathbb{Z}^+,
\end{equation}
where the period \( T \) represents the length of each stationary task, and the system adapts to the specific conditions of each task during this period.

In this work, the objective behind a given task is to minimize the long-term, time-averaged energy consumption while ensuring the queue stability by controlling the transmit power and the durations of transmissions and \gls{EH}. 
Towards this the goal of task $j$ is formalized as
\begin{subequations}\label{eqn:optimization_problem_1}
    \begin{eqnarray}
    \label{eqn:objective_function}
    \underset{\left(\PrimaryTransmissionPower, \EHTime, \PrimaryCommunicationTime, \SecondaryCommunicationTime\right)_t}{\text{minimize}}
    && \!\!\!\! \!\!\!\!
    \lim\limits_{T\to\infty} \frac{1}{T} \sum\limits_{t=0}^T \sum_{i=0}^1 \TransmissionPower \CommunicationTime, \\
    \label{eqn:time_constraint_1}
    \text{subject to}
    && \!\!\!\! \!\!\!\!
    0 \leq \EHTime, \PrimaryCommunicationTime, \SecondaryCommunicationTime \leq 1 \quad \forall{t}, \\
    \label{eqn:time_constraint_2} 
    && \!\!\!\! \!\!\!\!
    \textstyle \EHTime + \PrimaryCommunicationTime + \SecondaryCommunicationTime \leq 1 \quad \forall{t}, \\
    \label{eqn:queuing_stability_constraint}
    && \!\!\!\! \!\!\!\!
    \textstyle \lim_{t\to\infty} \frac{\bar{q}^j_{i,t}}{t} = 0 \quad i\in\{0,1\}, \\
    \label{eqn:Battery_capacity_constraint}
    && \!\!\!\! \!\!\!\!
    b^j_t \leq B \quad \forall{t}, \\
    \label{eqn:TX1_Power_Constraint}
    && \!\!\!\! \!\!\!\!
    \PrimaryTransmissionPower \leq \PrimaryMaxPower, \quad \forall{t}, \\
    \label{eqn:queue_dynamics_constraint}
    && \!\!\!\! \!\!\!\!
    \eqref{eqn:queue_dynamics}, \eqref{eqn:battery_queue_dynamics},
    \end{eqnarray}
\end{subequations} 
where $B$ is the maximum battery capacity. 
Constraints~\eqref{eqn:time_constraint_1}, ~\eqref{eqn:time_constraint_2} ensure feasible scheduling, ~\eqref{eqn:queuing_stability_constraint} ensures queue stability, while ~\eqref{eqn:Battery_capacity_constraint} and ~\eqref{eqn:TX1_Power_Constraint} satisfy energy availability for transmission.
The secondary transmitter’s power is fixed to reflect typical hardware constraints in low-power sensor nodes. Fixing the secondary transmitter’s power simplifies the problem by reducing control dimensionality, enabling more efficient optimization and learning. This design choice also reflects practical constraints in low-power sensor nodes, where hardware or protocol limitations often mandate fixed power levels to ensure reliable communication under tight energy budgets.

The overarching goal is to optimize the system’s decisions across all tasks in a non-stationary environment. 
A key drawback of traditional optimization methods is their slow adaptation to changing conditions in dynamic, non-stationary environments. 
This drawback motivates the development of a lifelong machine learning algorithm, which is designed to more effectively adapt the power and time control strategy to varying conditions. 
In the next section, we will present the \gls{RL} formulation, highlighting its limitations in these contexts. Building on this, we will introduce the \gls{L2RL} approach as a more adaptable solution tailored to handle these challenges.

\section{Leveraging \gls{L2RL} for Adaptive Multi-Task Optimization}\label{sec:L2RL_approach}

The optimization problem in~\eqref{eqn:optimization_problem_1} focuses on enhancing the \gls{WSN} performance in dynamic environments. 
Each environment can be treated as a stationary environment, represented by a characteristic tuple $\varphi^j = (\EHChannelScaleParameter, \PowerConversionEfficiency)$, where $\EHChannelNorm\sim f(\EHChannelScaleParameter)$, and $\PowerConversionEfficiency \in \mathscr{L}^j$.

Hence, a mechanism for effective knowledge transfer across environments is crucial. 
This calls for a lifelong learning framework, which supports leveraging knowledge from known tasks to new tasks, enabling efficient learning. 
Lifelong learning thus offers a compelling solution for navigating the evolving and uncertain nature of this problem. 

\subsection{Reinforcement Learning Formulation}

We model the decision-making problem in the dynamic \glspl{WSN} as a \gls{MDP}, formally defined as a tuple \(\langle \mathcal{S}, \mathcal{A}, \mathcal{P}, \mathcal{R}, \gamma \rangle\)~\cite{Sutton.1998}, where

\begin{itemize}
    \item \textbf{State space} \(\mathcal{S}\): Contains all possible states where a state \(\boldsymbol{s} = (q_0, q_1, b, h_0, h_1, \tilde{h}) \in \mathcal{S}\) represents the system status at any given time slot.

    \item \textbf{Action space} \(\mathcal{A}\): Contains all possible actions where an action \(\boldsymbol{a} = (p_0, \alpha_0, \alpha_1, \tilde{\alpha}) \in \mathcal{A}\) defines the decision variables.

    \item \textbf{Transition function} \(\mathcal{P}(\boldsymbol{s}'|\boldsymbol{s}, \boldsymbol{a})\): Defines the probability distribution over the next state \(\boldsymbol{s}'\), conditioned on the current state \(\boldsymbol{s}\) and action \(\boldsymbol{a}\). Transitions are governed by queue dynamics, battery updates, and stochastic channel variations.

    \item \textbf{Reward function} \(\mathcal{R}(\boldsymbol{s}, \boldsymbol{a})\): The reward function is defined as the negative of the objective~\eqref{eqn:optimization_problem_1}, penalizing more energy consumption, augmented with penalty terms to account for the constraints, satisfying queueing stability. The optimization problem~\eqref{eqn:optimization_problem_1} enforces hard constraints, while the \gls{RL} reward penalizes constraint violations, allowing temporary violations during learning. The reward function is given as
    \begin{equation}\label{eqn:reward_function_mdp}
    \mathcal{R}(\boldsymbol{s}^j_t, \boldsymbol{a}^j_t) = -\sum_{i=0}^1 \TransmissionPower \CommunicationTime - \nu\bigg((\BatteryStateDynamics - B)+\sum_{i=0}^1(\QueueStateDynamics-\TransmittedData)\bigg),
    \end{equation}
    where \(\nu\) is a penalty coefficient used to enforce queue stability and battery constraints.

    \item \textbf{Discount factor} \(\gamma \in [0,1]\): Models the trade-off between immediate and future rewards. We typically use \(\gamma \approx 1\) to prioritize long-term performance.
\end{itemize}

This MDP formulation provides a structured framework for applying \gls{RL} to learn adaptive control strategies under uncertain and time-varying conditions in \glspl{WSN}. During the agent’s interaction with the environment, a trajectory \(\tau\) is collected, representing the sequence of interactions over time.  
The trajectory $\tau = \{(\boldsymbol{s}_t^j, \boldsymbol{a}_t^j, \mathcal{R}(\boldsymbol{s}^j_t, \boldsymbol{a}^j_t))\}_{t=0}^T$ captures the full interaction history of the agent and is used to infer information about the task. 
The agent’s decision-making process is governed by a policy \(\pi_{\boldsymbol{\theta}^j}(\boldsymbol{a}_t^j | \boldsymbol{s}^j_t)\), which denotes the probability of selecting action $\boldsymbol{a}_t^j$ given state $\boldsymbol{s}^j_t$ and policy parameters \(\boldsymbol{\theta}^j \in \mathbb{R}^N\), where $N$ is the number of variables in the state space, and 
$\Pi = \left\{ \pi_{\boldsymbol{\theta}^j} \;\middle|\; \boldsymbol{\theta}^j \in \mathbb{R}^N \right\}$ denotes the policy space.
The agent learns and updates \(\pi_{\boldsymbol{\theta}^j}(\boldsymbol{a}_t^j | \boldsymbol{s}^j_t)\) based on accumulated experience, adapting to the dynamics of the specific task~\cite{Chen.18}. 

Since the agent does not have prior knowledge of the task specification. i.e., the environment's transition dynamics and reward structure, it must infer the task by analyzing trajectories collected through interactions. 
This process involves extracting statistical patterns and dynamic characteristics from $\tau$ to approximate the latent task configuration, which encompasses the hidden characteristics of the environment.
The likelihood of observing a trajectory $\tau$ under policy $\pi_{\boldsymbol{\theta}^j}$ is given by the distribution
\begin{equation}\label{eqn:probability_distribution_trajectory}
    p_{\boldsymbol{\theta}^j}(\tau) = C_0(s_0) \prod\nolimits_{t=0}^{T} p(s_{t+1}^j | s_t^j, \boldsymbol{a}_t^j) \pi_{\boldsymbol{\theta}^j}(\boldsymbol{a}_t^j | s_t^j),
\end{equation}
where $C_0(s_0)$ is the initial state distribution.
Following that, the expected return for task $j$ under the policy parametrized by $\boldsymbol{\theta}^j$ is defined as
\begin{equation}\label{eqn:expected_return_for_task_j}
     \Gamma(\boldsymbol{\theta}^j) = \int p_{\boldsymbol{\theta}^j}(\tau) \rho(\tau) d\tau,
\end{equation}
where $\rho(\tau)=\frac{1}{T}\sum_{t=0}^T \mathcal{R}(s^j_t, a^j_t)$ is the gain for trajectory $\tau$. 
By integrating over all possible trajectories $\tau$ weighted by their likelihood $p_{\boldsymbol{\theta}^j}(\tau)$ under the policy $\pi_{\boldsymbol{\theta}^j}$, the expected return $\Gamma(\boldsymbol{\theta}^j)$ quantifies the overall performance of the policy for task~$j$.

Traditional \gls{RL} assumes stationary environments, slowing the rate of adaptability across changing conditions. In dynamic environments, \gls{RL} struggles with inferring unknown task dynamics.
\gls{L2RL} addresses these limitations by enabling the agent to learn sequentially, adapt to new environments, and leverage prior knowledge for improved performance across changing tasks.

\subsection{Lifelong Learning Framework}

In dynamic wireless sensor environments, new tasks arrive over time, reflecting the evolving nature of external conditions and application demands. 
To address this, we consider a lifelong learning framework where the agent incrementally learns from each task. At any time, the agent has access to a set of known tasks it has encountered and anticipates the arrival of new tasks. 
The goal is to systematically extract knowledge from past tasks and update the \gls{KB}, a continually updated repository of learned information, to efficiently adapt to future tasks.

Each task $j$ is modeled as a \gls{MDP}, and the long-term goal is to maximize the average expected return across all encountered tasks as\begin{equation}\label{eqn:maximizing_expected_average_return_all_tasks}
    \underset{\Pi}{\text{max}} \quad \lim_{M \to \infty} \frac{1}{M} \sum\nolimits_{j=1}^M \Gamma(\boldsymbol{\theta}^j).
\end{equation}
To facilitate knowledge transfer, we decompose the policy parameter for each task as $\boldsymbol{\theta}^j = \mathbf{G}\boldsymbol{v}^j$, where $\mathbf{G} \in \mathbb{R}^{N \times Z}$ is a shared latent basis representing the \gls{KB}, and $\boldsymbol{v}^j \in \mathbb{R}^Z$ is a sparse task-specific coefficient vector. 
This factorization enables generalizable knowledge to be captured in $\mathbf{G}$ while adapting to each task via $\boldsymbol{v}^j$ \cite{BouAmmar.15}.

To extract robust and interpretable representations, we define the lifelong learning loss as~\cite{BouAmmar.15}
\begin{equation}\label{eqn:lifelong_learning_loss_function}
    J(\mathbf{G}) = \frac{1}{M} \sum_{j=1}^M \min_{\boldsymbol{v}^j} \left[ -\Gamma(\boldsymbol{\theta}^j) + \mu_1 \|\boldsymbol{v}^j\|_1 \right] + \mu_2 \|\mathbf{G}\|_F^2,
\end{equation}
where $\mu_1$ promotes sparsity in $\boldsymbol{v}^j$, $||.||_F$ denotes Frobenius norm, and $\mu_2$ regularizes the complexity of the \gls{KB}. However, optimizing this loss function across all tasks is impractical during training, as new tasks continually arrive and the agent may not access all task data simultaneously. To address this, we approximate $-\Gamma(\boldsymbol{\theta}^j)$ using a second-order Taylor expansion around the optimal policy parameter $\boldsymbol{\beta}^j$, where $\nabla_{\boldsymbol{\theta}^j}(-\Gamma(\boldsymbol{\theta}^j)) = 0$. This leads to an incremental formulation given by
\begin{equation}\label{eqn:approximation_$J(G)$}
    \hat{J}(\mathbf{G}) = \frac{1}{M} \sum_{j=1}^M \min_{\boldsymbol{v}^j} \left[ \|\boldsymbol{\beta}^j - \mathbf{G} \boldsymbol{v}^j \|_{\boldsymbol{Q}^j}^2 + \mu_1 \|\boldsymbol{v}^j\|_1 \right] + \mu_2 \|\mathbf{G}\|_F^2,
\end{equation}
where $\boldsymbol{Q}^j$ is the Hessian matrix capturing curvature information. 

To enable scalable and adaptive policy learning across a stream of tasks, we adopt a lifelong learning framework that incrementally refines a shared latent \gls{KB}. 
Rather than recomputing the base from scratch for each task, we integrate new information using a recursive structure with exponential moving averages of sufficient statistics.
For each new task $j$, we first estimate the task-optimal policy parameters $\boldsymbol{\beta}^j$ and curvature matrix $\boldsymbol{Q}^j$ using the policy gradient \gls{RL} framework. 
Then, we compute the sparse task-specific coefficient vector $\boldsymbol{v}^j$ as follows
\begin{equation}\label{eqn:task_adaptation}
\boldsymbol{v}^j \leftarrow \arg \min_{\boldsymbol{v}} \left( \| \boldsymbol{\beta}^j - \mathbf{G} \boldsymbol{v} \|_{\boldsymbol{Q}^j}^2 + \mu_1 \| \boldsymbol{v} \|_1 \right).
\end{equation}
Using which, two task specific matrices are computed as
\begin{equation}\label{eqn:task_running}
    \mathbf{X}^j = \boldsymbol{v}^j (\boldsymbol{v}^j)^\top 
    \qquad \text{and} \qquad
    \mathbf{Y}^j = \boldsymbol{\beta}^j (\boldsymbol{v}^j)^\top.
\end{equation}
Then, the running estimates of cumulative task statistics $\mathbf{X} \in \mathbb{R}^{Z \times Z}$ and $\mathbf{Y} \in \mathbb{R}^{N \times Z}$ are given by,
\begin{align}
\label{eqn:X_running}
\mathbf{X} &\leftarrow (1 - \eta) \mathbf{X} + \eta \mathbf{X}^j, \\
\label{eqn:Y_running}
\mathbf{Y} &\leftarrow (1 - \eta) \mathbf{Y} + \eta \mathbf{Y}^j.
\end{align}
%
where $\eta \in (0, 1)$ is a learning rate controlling the influence of the current task. 
With these statistics, the \gls{KB} is updated as follows:
\begin{equation}\label{eqn:G_update}
\mathbf{G} \leftarrow \mathbf{Y} \mathbf{X}^\dagger,
\end{equation}
where $\mathbf{X}^\dagger$ denotes the Moore–Penrose pseudoinverse, which is a robust inversion even with non full-rank matrices \cite{ben.2003}.  
This yields the minimum-norm solution to the least-squares problem $\min_{\mathbf{G}} \| \mathbf{Y} - \mathbf{G} \mathbf{X} \|_F^2$, ensuring stable \gls{KB} updates even under sparse or ill-conditioned input statistics.
This formulation enables continual integration of new task knowledge while retaining and compressing prior experience in a bounded memory footprint.


The proposed \gls{MT-L2RL} algorithm implements the above framework for scalable and adaptive lifelong learning. 
For each incoming task, it first estimates the task-optimal policy using \gls{RL}, then computes a sparse representation over the shared latent basis. 
The algorithm incrementally updates cumulative statistics using exponential moving averages, enabling the \gls{KB}, $\mathbf{G}$, to evolve over time without storing all prior task data. 
This process supports continual policy refinement and transfer by efficiently integrating new task-specific information into a compact and reusable representation.
The algorithm is detailed in Algorithm~\ref{alg:Algortihm_1}.
\begin{algorithm}[t]
\caption{\gls{MT-L2RL}: Multi-Task Lifelong Reinforcement Learning with Incremental Knowledge Update}
\label{alg:Algortihm_1}
\begin{algorithmic}[1]
\STATE \textbf{Input:} 
\begin{itemize}
    \item Initialize latent \gls{KB}, \( \mathbf{G} \in \mathbb{R}^{N \times Z} \),
    \item Initialize cumulative statistics \( \mathbf{X} \in \mathbb{R}^{Z \times Z}, \mathbf{Y} \in \mathbb{R}^{N \times Z} \)
\end{itemize}
\STATE \textbf{Hyperparameters:}
\begin{itemize}
    \item Sparsity regularization \( \mu_1 \), \gls{KB} regularization \( \mu_2 \)
    \item Learning rate \( \eta \in (0,1) \)
    \item Number of tasks \( M \)
\end{itemize}

\FOR{each incoming task \( j = 1, \dots, M \)}
    \STATE \textbf{1. Estimate Task-Specific Parameters:}
    \STATE \quad Use RL to estimate optimal policy \( \boldsymbol{\beta}^j \in \mathbb{R}^N \) and curvature matrix \( \boldsymbol{Q}^j \in \mathbb{R}^{N \times N} \)
    
    \STATE \textbf{2. Solve for Sparse Task Encoding~\eqref{eqn:task_adaptation}}

    \STATE \textbf{3. Compute Task Statistics~\eqref{eqn:task_running}}

    \STATE \textbf{4. Update Cumulative Statistics~\eqref{eqn:X_running}}

    \STATE \textbf{5. Update \gls{KB}~\eqref{eqn:G_update}}
\ENDFOR

\STATE \textbf{Output:} Final \gls{KB} \( \mathbf{G} \) and task encodings \( \{ \boldsymbol{v}^j \}_{j=1}^M \)
\end{algorithmic}
\end{algorithm}

\section{Simulation Results and Analysis} \label{sec:simulation_results}

The simulation is conducted in a dynamic \gls{WSN} environment where tasks arrive sequentially over time. Each task represents a distinct wireless communication scenario characterized by varying environmental parameters, such as fading condition and power conversion efficiency. These tasks simulate the evolution of the network dynamics, which requires the agent to adapt its policy accordingly.

The system operates on a bandwidth of $W = \SI{5}{\mega\hertz}$ with a noise power spectral density of $N_0 = \SI{-120}{\dBm}$. 
To reflect moderate traffic conditions, the density of arrival data is set to be $\SI{1}{\kilo\bit\per\second}$.
Transmission power levels are limited to $\PrimaryMaxPower = \SI{0.03}{\watt}$ and $\SecondaryMaxPower = \SI{0.01}{\watt}$, and $B = \SI{5}{\joule}$. 
The main hyperparameters used throughout the experiments are given as following. 
$\gamma=0.99$ to emphasize long-term rewards.
The regularization parameters are set to $\mu_1 = 0.01$ and $\mu_2 = 0.01$, based on preliminary tuning for stability and generalization. 
Each task operates over a stationary period of length $T = 500$ time slots, during which the environmental conditions are held constant.

For policy optimization in each task, we adopt a standard policy gradient \gls{RL} algorithm as the base learner. 
The algorithm updates the policy parameters $\boldsymbol{\theta}^j$ to maximize the expected return $\Gamma(\boldsymbol{\theta}^j)$ as defined in~\eqref{eqn:expected_return_for_task_j}. While Rayleigh fading is a common choice, non-stationarity is enforced by varying the distribution parameters and system-level settings across tasks. In addition, the power conversion efficiency $\PowerConversionEfficiency \in \mathcal{L}^j$ is sampled from different ranges between tasks. This modeling introduces the need for continual policy adaptation.


\begin{table}[t]
\centering
\caption{\Gls{EH} conditions for Training and Testing Tasks}
\label{tab:sim_parameters}
\begin{tabular}{ccc}
\toprule
\textbf{Parameter} & \textbf{Training (25 tasks)} & \textbf{Testing (4 tasks)} \\
\midrule
\(\EHChannelScaleParameter\) & \(\mathcal{U}(0.5, 1.5)\) & \{0.6, 1.0, 1.4, 1.8\} \\
\(\PowerConversionEfficiency\) & \(\mathcal{U}(0.3, 0.6)\) & \{0.35, 0.45, 0.55, 0.65\} \\
\bottomrule
\end{tabular}
\end{table}

The training phase consists of sequential task arrivals where the agent updates both the \gls{KB} and task-specific representations. In contrast, during the testing phase, four new tasks are presented without further updates to the \gls{KB}; only fast adaptation of $\boldsymbol{v}^j$ is performed. This setup enables us to evaluate the adaptability of the proposed \gls{MT-L2RL} framework to previously unseen tasks. The specific parameter ranges that differentiate the training and testing phases are summarized in Table~\ref{tab:sim_parameters}. We compare our \gls{MT-L2RL} method with the following two baselines: i) \textbf{Standard RL:} A vanilla reinforcement learning agent trained from scratch for each task independently and 
ii) \textbf{Lyapunov Optimization-based Method:} A task-specific solver without learning from prior tasks.
%


\begin{figure}[t]
    \centering
    \includegraphics[width=8.8cm]{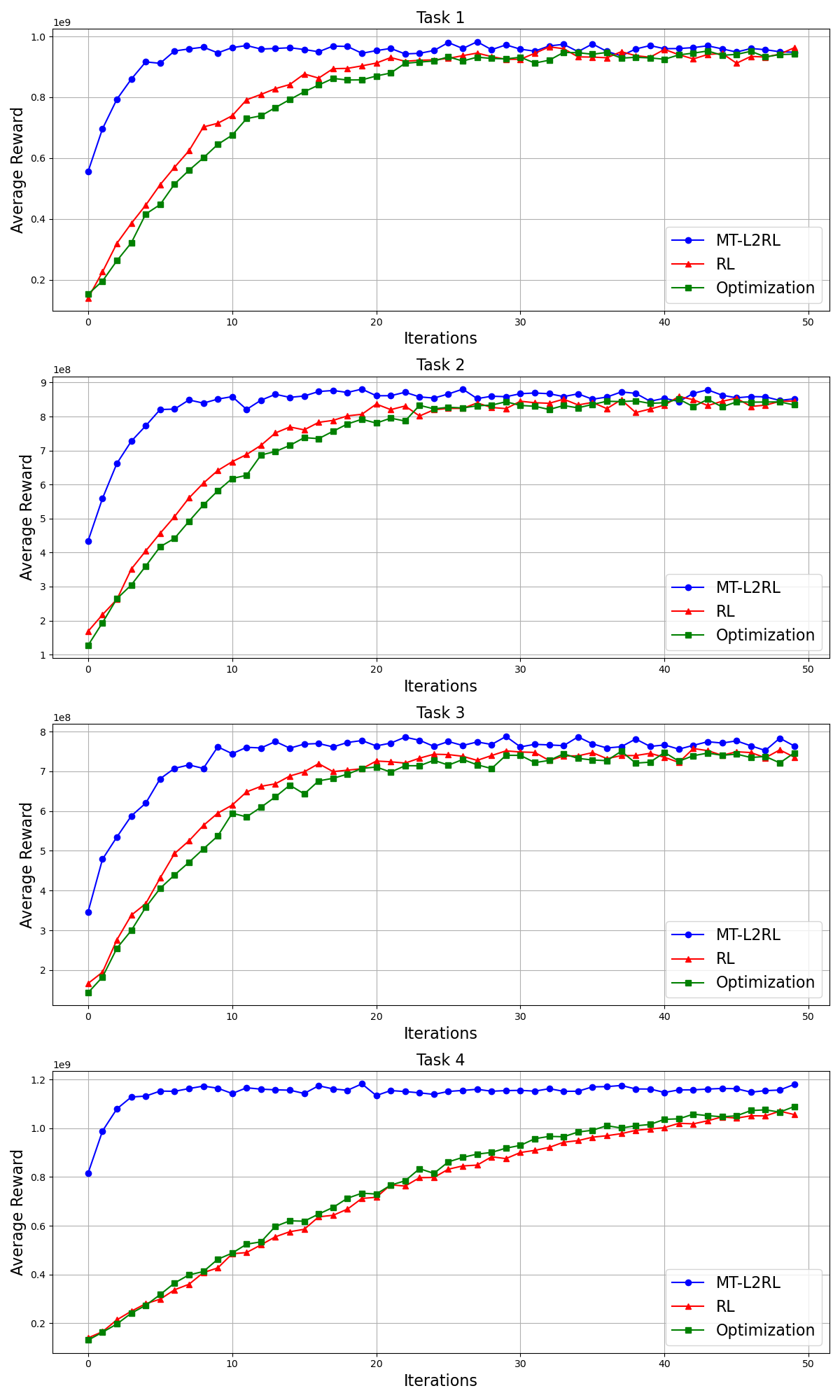}
    \caption{Performance comparison on new tasks.}
    \label{fig: simulation-1}
\end{figure}  


Figure~\ref{fig: simulation-1} illustrates the learning performance on newly arriving tasks during the testing phase. 
The term iterations refers to the number of policy update steps the agent performs during adaptation to a test task. 
Each iteration involves interaction with the environment, collection of feedback, and update of the task-specific parameters, while keeping the shared \gls{KB} fixed.
The agent must adapt quickly using the knowledge accumulated across earlier tasks. 
Our \gls{MT-L2RL} model consistently demonstrates faster convergence compared to both baseline models. Specifically, it achieves up to 60\% faster convergence in favorable conditions and at least 30\% faster in more challenging tasks.
Unlike the baselines, having cold-start, \gls{MT-L2RL} initializes its task-specific parameters using prior knowledge encoded in the shared base \( \mathbf{G} \), leading to significant acceleration in learning. As shown in~Fig.\ref{fig: simulation-1}, our proposed solution adapts significantly faster, highlighting the strength of lifelong learning for dynamic environments.

This experimental setup validates the key claim of our framework: rather than excelling on known tasks, the true strength of \gls{MT-L2RL} lies in its ability to rapidly generalize to new, unseen tasks by leveraging accumulated experience. This aligns with the broader motivation behind lifelong learning—quick adaptation, rather than static optimality.

\section{Conclusion}\label{sec:conclusion}
This paper proposes a \gls{L2RL} approach to optimize the long-term time-averaged energy consumption in an \gls{EH} wireless sensor network over non-stationary environments, while maintaining queuing stability. 
Our proposed solution leverages knowledge transfer between tasks to avoid catastrophic forgetting. 
This procedure enables the agent to adapt its policies to unknown environments. 
The proposed solution is compared to two benchmarks, the Lyapunov framework and the \gls{RL} setting. 
Simulation results show that the proposed algorithm can achieve up to $60\%$ progress in convergence compared to the benchmarks. 
A promising future direction involves extending the current framework to distributed or multi-agent \gls{RL} paradigms. 
These extensions could allow each sensor node or cluster to operate as an autonomous agent, collaborating or competing with others to optimize energy and communication efficiency while maintaining global network objectives.

\bibliographystyle{IEEEtran}
\bibliography{References}

\end{document}